\documentclass[aps,prx,twocolumn,showpacs,superscriptaddress]{revtex4-1}
\usepackage{latexsym}
\usepackage{amssymb}
\usepackage{graphicx}
\usepackage{amsmath}
\usepackage{bm}
\usepackage[colorlinks,
          linkcolor=black,
            citecolor=black,
            urlcolor=blue
           ]{hyperref}
\usepackage{verbatim}
\usepackage{mathrsfs}
\usepackage{extarrows}
\usepackage{comment}

\begin{document}

\title{Symmetry and microscopic constraints on Hall conductances and a Wiedemann-Franz law: a view from the interface}

\author{Yuan Yao}
\email{smartyao@issp.u-tokyo.ac.jp}
\affiliation{Institute for Solid State Physics, University of Tokyo, Kashiwa, Chiba 277-8581, Japan}
\affiliation{Kavli Institute for Theoretical Physics, University of California, Santa Barbara, California 93106, USA}

\begin{abstract}
We introduce an unconventional regularization of condensed matter effective field theory by a non-local bulk topological regulator in one higher dimension, which is equivalent to a bulk-interface-bulk formulation. We show its necessity motivated by a (0+1)-dimensional system. Although the system can be strictly defined on the lattice in its own dimensions, the bulk regulator is important to correctly derive its physical observables, which cannot be captured by any local regulator. We explicitly obtain a nontrivial constraint on the thermal and electric integer Hall conductances of half-filled translation invariant $N$-flavor gapped fermionic system on a two-dimensional lattice possessing a unique ground state with uniform rational magnetic fluxes per unit cell in the presence of the onsite $U(N)$ symmetry. The Wiedemann-Franz law is shown to be obeyed by the Hall conductances regardless of arbitrarily strong interactions. We further obtain several no-go theorems of the possible symmetric gapped phases. 
\end{abstract}

\maketitle

%%%%%%%%%%%%%%%%%%%%%%%%%%%%%%%%%%%%%%%%%%%%%%%%
%%%%%%%%%%%%%%%%%%%%%%%%%%%%%%%%%%%%%%%%%%%%%%%%
%%%%%%%%%%%%%%%%%%%%%%%%%%%%%%%%%%%%%%%%%%%%%%%%
\section{Introduction}
Regularizations are the processes by which one renders divergent quantities finite in quantum field theories. Different regularization schemes were believed to give the same results on physical observables defined on underlying lattices when the perturbation theories are done to all orders. 
However, this folklore is not generally true and its failure is attributed to a new degree of freedom to regularize field theories by a topological term defined in one higher dimensional bulks, which has its origin in the field-theoretical description of the boundary states of symmetry-protected topological (SPT) phases~\cite{Gu:2009aa,Chen:2013aa,Pollmann:2012aa}. The boundary theories are anomalous in the sense that they cannot be realized in their own dimensions respecting certain onsite symmetries without being attached to an SPT bulk~\cite{Ryu:2012ac, Wen:2013aa, Wang:2013aa, Hsieh:2014aa, Hsieh:2016aa, Seiberg:2016aa, Witten:2016ab, Witten:2019aa}. 
%As an essential consequence, the boundary states cannot be symmetrically gapped without intrinsic topological orders and the theory is well-defined only when attached to the SPT bulk. 
Thus there is no canonical separation of the boundary and its bulk, and the partition function of the boundary theory must also include a bulk term~\cite{Seiberg:2016aa,Witten:2019aa}, which will be called ``bulk regularization''. 

We will show that bulk regularizations also exist in quantum field theories realizable in their own dimension(s) and have observable consequences: they are essential to obtain correct electric and thermal Hall conductances from field theories, which are useful quantities to distinguish various quantum phases. 
Although it was argued by earlier works~\cite{Hatsugai:1996aa,Oshikawa:1994aa,Watanabe:2010aa,Bernevig:2013aa} that Hall conductances cannot be fully captured by low-energy field theories due to ``spectator'' fermion contribution from higher momenta, we will see that such spectator contributions are actually non-perturbatively controllable in the presence of certain symmetries. 
The topological nature of bulk regularizations and symmetries will enable us to analyze electric and thermal Hall conductances in strongly-correlated systems in a unified way by concise low-energy effective theories. 
%in contrast to the situation that Hall conductances are generically difficult to calculate even in non-interacting band insulators by Chern numbers~\cite{Thouless:1982aa,Kohmoto1985}. 

%The thermal and electric Hall conductances will be also related and treated in a unified way by the field-theoretical method.
The Wiedemann-Franz (W-F) law~\cite{Franz:1853aa,Chester:1961aa} for non-interacting integer quantum Hall (IQH) phases~\cite{Kane:1997aa,Wang:2014aa,Wang:2014ab,Wang:2014ac} states that, in \emph{free} fermion systems gapped with a unique ground state, the ratio of thermal Hall conductance $\kappa_{xy}$ and electric Hall conductance $\sigma_{xy}$: $\kappa_{xy}/\sigma_{xy}=LT=1$ in the low-temperature limit, where we scale the Lorenz number $L$ to be inverse of the temperature $T$: $L=\pi^2/3\equiv T^{-1}$. 
In the presence of strong interactions in the gapped system with a unique ground state, the law is generically broken, but $\kappa_{xy}=\sigma_{xy}\mod8$ instead~\footnote{It can also be directly seen by the fact that either the gravitational or electromagnetic Chern-Simons term alone can be well-defined on any spin$_c$ manifold only if its level is divisible by eight~\cite{Seiberg:2016aa}.}, where the mod-$8$ ambiguity can be understood by the formation of cluston IQH states by binding three electrons in a Chern band or a fermionic construction of unfractionalized bosonic phases~\cite{Lu:2012aa,Wang:2014aa,Wang:2014ab,Wang:2014ac}. 
Therefore, the conditions, under which the IQH W-F law can be restored in strongly-correlated systems, are useful in real measurements. 

In this work, we show that the mod-$8$ uncertainty above is eliminated --- the IQH W-F law $\kappa_{xy}/\sigma_{xy}=1$ is strictly respected and there is a nontrivial constraint $\kappa_{xy}=\sigma_{xy}=Nq/2\mod Nq$, if an onsite $U(N)$ symmetry and a magnetic translational symmetry ${\cal{M}}$~\cite{Brown:1964aa,Zak:1964aa,Zak:1964ab,Florek:1996aa,Lu:2017aa} are preserved in arbitrarily interacting gapped systems with a unique ground state, constituted by half-filled fermions on a two-dimensional lattice with rational magnetic fluxes $2\pi(p/q)$ per unit cell, where the integers $p$ and $q$ coprime. 
The $U(N)$ symmetry is useful and realizable in ultracold atoms on optical lattices to study $SU(N)$ generalized spin systems at low energy~\cite{Affleck:1987aa,Affleck:1988aa,Wu:2003aa,Honerkamp:2004aa,Cazalilla:2009aa,Gorshkov:2010aa,Taie:2012aa,Pagano:2014aa,Scazza:2014aa,Zhang:2014aa} and ${\cal M}$ is a physical generalization of conventional lattice translations when a uniform external magnetic field is present~\cite{Brown:1964aa,Zak:1964aa,Zak:1964ab,Florek:1996aa,Lu:2017aa}. 
%Specifically, such Hall conductances are always non-vanishing by the symmetry requirement above. 
The result implies that measurements on $\kappa_{xy}$ can be done by measuring $\sigma_{xy}$ instead and we can obtain several no-go theorems on the possible symmetric insulating phases, which are consistent with generalized Lieb-Schultz-Mattis (LSM) theorems~\cite{Lieb:1961aa,Oshikawa:2000aa, Hastings:2004ab, Affleck:1986aa, Cheng:2016aa,Yao:2019aa,Yao:2019ab}. 
%These are direct consequences of the bulk regularization, which plays an essential role in deriving the electric and thermal responses. 

\section{Motivation by $(0+1)$ dimension}
We motivate the bulk regularization by the following system: a $(0+1)$-dimensional system has a single spin-1/2 with degenerate spin-up and down states as its low-energy theory and is assumed to possess a global spin-rotation symmetry on the lattice at this gapless point~\footnote{Actually, we do not have to require the full spin-rotation symmetry at the gapless point: only spin-rotation around $z$-axis $U(1)_z$ and $\pi$-rotation around $y$-axis ${\cal{C}}$ are needed.}. 
Let us consider the following question: if the system is gapped with a unique ground state by strong interactions respecting the spin-rotation symmetry around $z$-axis $U(1)_z$, what $\hat{S}_z$-eigenvalues can the unique ground state have? 

We first start from the original gapless point possessing $U(1)_z\rtimes{\cal{C}}$ by the assumption, where ${\cal{C}}$ denotes the spin $\pi$-rotation around $y$-axis. Then we describe it by a quantum field theory in the imaginary time $\tau=it$ on a circle $X_1=S^1$: 
\begin{eqnarray}
L_0[\psi,{A}_\tau]=\psi^\dagger(\partial_\tau-iA_\tau)\psi, 
\end{eqnarray}
where we have defined a fermionic operator $\{\psi,\psi^\dagger\}=1$ and $\{\psi,\psi\}=0$ with the mapping $\hat{S}_x=(\psi^\dagger+\psi)/2$, $\hat{S}_y=i(\psi^\dagger-\psi)/2$ and $\hat{S}_z=\psi^\dagger\psi-1/2$ for the low-energy spin-1/2 and $A_\tau$ is the background $U(1)_z$ gauge field since the spin rotation by $\alpha$-angle around $z$-axis is represented by $\psi\rightarrow\psi\exp(i\alpha)$, and ${\cal{C}}$ by $\psi\rightarrow\psi^\dagger$ and $A_\tau\rightarrow-A_\tau$. 
Thus the charge coupled to $A_\tau$-field is exactly $S_z$. 

Then we add a local $U(1)_z$-respecting interaction $L_\text{int}[\psi]$ to gap this system and a plausible Lagrangian regularized by a Pauli-Villars (P-V) regulator can be $L_\text{P-V}[\psi,\chi,A_\tau]=L_0[\psi,A_\tau]+L_\text{int}[\psi]+\chi^\dagger(\partial_\tau-iA_\tau+\mu^{\,}_\text{P-V})\chi$, where $\chi$ is a (dynamical) P-V bosonic spinor with a mass $\mu^{\,}_\text{P-V}$ as a momentum cut-off. 
Indeed, $L_\text{P-V}$ satisfies the $U(1)_z$-symmetry requirement. 
The partition function is denoted as $z^{\,}_\text{P-V}[A_\tau]$. 
After the matter fields $\psi$ and $\chi$ are integrated out, the most general form of the partition function is, modulo a non-universal coefficient, $z^{\,}_\text{P-V}[A_\tau]\propto\exp\left(i\int_{X_1} d\tau n_0A_\tau\right)$ where $n_0\in\mathbb{Z}$~\cite{Dirac:1931aa}. 
It implies that $S_z=n_0\in\mathbb{Z}$. 
This conclusion is actually incorrect because we can simply take the lattice system to be a single spin-1/2, which has a half-integer $S_z=\pm1/2$ rather than integers when gapped by magnetic fields along $z$-axis. 

The problem of the ``derivation'' above can be seen as follows. Let us switch off the interaction by $L_\text{int}[\psi]=0$ in $L_\text{P-V}[\psi,\chi,A_\tau]$ with a non-interacting partition function $z^{\,}_{0,\text{P-V}}[A_\tau]$ and we have a free spin-1/2 Hamiltonian $i\hat{A}_\tau\hat{S}_z$ at low energy enjoying a ${\cal{C}}$ symmetry at the lattice scale.
However, the field theory $z^{\,}_{0,\text{P-V}}[\psi,\chi,A_\tau]$ breaks the ${\cal{C}}$ symmetry due to the mass term of P-V regulator. 
Indeed, no \emph{local} regularization preserving $U(1)_z\rtimes{\cal{C}}$ symmetry exists because of a mixed $U(1)_z\rtimes{\cal{C}}$ anomaly~\cite{Elitzur:1986aa}. 
Thus there is a symmetry mismatch between the effective theory $z^{\,}_{0,\text{P-V}}[A_\tau]$ and the underlying lattice model at the gapless point. 
The only symmetry-consistent regularization for this gapless point is the following (non-local) bulk regularization of the partition function, which cannot be captured by any local regularization~\footnote{We have chosen, without loss of generality, $\mu^{\,}_\text{P-V}>0$ so that there is no sign redundancy of the theta-term.}: 
\begin{eqnarray}
\label{bulk_0}
z_0[A_\tau]=z^{\,}_{0,\text{P-V}}[A_\tau]\exp\left(\int_{Y_2} i\pi\frac{dA}{2\pi}\right),
\end{eqnarray}
which respects $U(1)_z\rtimes{\cal{C}}$ at the gapless point in the price of an auxiliary bulk $Y_2$ whose boundary $\partial Y_2$ is our system $X_1$ in $(0+1)$ dimension. 
To see its ${\cal{C}}$ invariance, we note that the phase of $z_0[A_\tau]$ is a real number $(-1)^{{\cal{I}}_{2D}}$ by the Atiyah-Patodi-Singer (APS) index theorem, where the integer ${\cal{I}}_{2D}$ is the Dirac index on $Y_2$~\cite{Atiyah:1963aa,Atiyah:1968aa,Atiyah:1975aa}. 
Furthermore, the bulk-regularized partition function (\ref{bulk_0}) can be shown to be equivalent, modulo a non-universal factor, with the partition function of a bulk Dirac fermion defined on a manifold $\tilde{Y}_2$ by pasting $Y_2$ and the vacuum $Y_{2,\text{vac}}$ along their interface $X_1=\partial Y_2=-\partial Y_{2,\text{vac}}$. 
The bulk Dirac mass, deeply in $Y_2$, has an opposite sign to the bulk P-V regulator mass and the same sign as the bulk regulator mass deeply in $Y_{2,\text{vac}}$, forming a bulk-interface-bulk picture~\cite{Jackiw:1976aa,Suppl}. 
The gapless system $L_0[\psi,A_\tau]$ can be thought to lie on the interface where the bulk Dirac mass vanishes. 
Then we switch on the interaction $L_\text{int}[\psi]\neq0$ to gap the system so that it has a unique ground state. 
Although $L_\text{int}[\psi]$ inevitably breaks ${\cal{C}}$ symmetry, it is still a local interaction on $(0+1)$-dimensional $X_1$. 
This locality is translated as follows in the bulk-interface-bulk formulation: the interaction $L_\text{int}[\psi]$ only acts near the interface $X_1$ and thus the deep bulks on its both sides remain unchanged. Thus, after the matter fields on $\tilde{Y}_2$ are integrated out, we calculate the resultant interacting partition function as~\cite{Suppl}
\begin{eqnarray}
\label{partition_function_0}
z[A_\tau]&=&\exp\left(\int_{\tilde{Y}_2} i\theta_0(x)\frac{dA}{2\pi}\right)\nonumber\\
&=&\exp\left[i(n_0+1/2)\int_{X_1} A_\tau\right], 
\end{eqnarray}
where the integration is done by part with $\theta_0(x)=-2\pi n_0$ deeply in $Y_{2,\text{vac}}$ ($n_0\in\mathbb{Z}$) and $\theta_0(x)=\pi$ deeply in $Y_2$~\cite{theta}. 
This result can be also obtained by the fact that the vacuum theta-angle is identified with $2\pi\mathbb{Z}$ due to the quantization $\int_{\tilde{Y}_2}dA/2\pi\in\mathbb{Z}$ for any orientable closed $\tilde{Y}_2$~\cite{Dirac:1931aa}.
Equation (\ref{partition_function_0}) is also physically sensible since its last line is independent of the auxiliary extension $Y_2$. 
Thus we obtain the correct $S_z=n_0+1/2\in\mathbb{Z}+1/2$, where the essential ``$1/2$'' results from the bulk regularization in (\ref{bulk_0}) and it cannot be captured by any local $U(1)_z$-symmetric regulator in (0+1) dimension: the P-V regulator can only contribute to the integer $n_0$. 

%Physically, we can alternatively first view the system at the gapless point, where the spin-up and down states are degenerate, as a boundary state of the Haldane phase protected by $U(1)_z\rtimes{\cal{C}}$ and the bulk term corresponds to the partition function of the bulk Hamiltonian, e.g. the Affleck-Kennedy-Lieb-Tasaki model~\cite{Affleck:2004aa}. 
%Then we introduce magnetic field along $z$-axis localized at the boundary, which only penetrates the bulk by a finite distance. 
%The integer $n_0$ in (\ref{partition_function_0}) is exactly the number of spin-$1$(s) flipped by the boundary magnetic field. 

\section{Translation invariant fermionic systems in $(2+1)$ dimensions with magnetic fields}
The minimal coupling $S_zA_\tau$ in the toy model above can be considered as a Chern-Simons (C-S) coupling in $(0+1)$ dimension. 
Therefore, higher dimensional C-S couplings are also expected to be restricted by symmetries.
Such restrictions constrain the permitted integer electric and thermal Hall conductances in $(2+1)$ dimensions as we will see below. 

Let us consider a half-filled square lattice of $N$-flavor fermions and we apply a uniform external magnetic field with rational fluxes $2\pi(p/q)$ per plaquette, where the integers $p$ and $q$ coprime and $q>0$ is assumed without loss of generality. 
Although the system is physically translationally invariant along two axises, the conventional lattice translations are broken by a gauge fixing. 
A more physical translation in our interest is the magnetic translations ${\cal{M}}=\{{\cal{M}}_{x},{\cal{M}}_{y}\}$, a combination of lattice translations and gauge transformations~\cite{Zak:1964aa,Zak:1964ab,Lu:2017aa}. 
Their commutator
\begin{eqnarray}
\label{inv}
{\cal{M}}_x{\cal{M}}_y{\cal{M}}_x^{-1}{\cal{M}}_y^{-1}=\exp\left[-i2\pi(p/q)\hat{F}\right], 
\end{eqnarray}
where $\hat{F}$ is the fermion number operator, is gauge invariant. 
Under a certain gauge fixing by which the lattice translation along $x$-axis is broken down to a $q$-lattice translation, ${\cal{M}}$ is represented by
\begin{eqnarray}
\label{M}
&&{\cal{M}}_x:\,c_{\vec{r}}\rightarrow c_{\vec{r}+\hat{x}}\exp\left(-i2\pi\frac{p}{q}r_y\right);\nonumber\\
&&{\cal{M}}_y:\,c_{\vec{r}}\rightarrow c_{\vec{r}+\hat{y}}, 
\end{eqnarray}
where the $N$-flavor fermion annihilation operator $c_{\vec{r}}=\left[c^{(1)}_{\vec{r}},\cdots,c^{(N)}_{\vec{r}}\right]$ is defined at the lattice site $\vec{r}=(r_x,r_y)$ with $\hat{x}$ and $\hat{y}$ as unit vectors. 
We are interested in the constraint on the (integer) electric and thermal Hall conductances of this half-filled system when it is gapped with a unique ground state in the presence of $U(N)$ symmetry and the magnetic translational symmetry ${\cal{M}}$. 
Additionally, the generalized LSM theorems~\cite{Lieb:1961aa,Oshikawa:2000aa, Hastings:2004ab, Affleck:1986aa, Cheng:2016aa,Yao:2019aa,Yao:2019ab} require $q=2N_c$ for some integer $N_c$ so that the system can be gapped with a symmetric unique ground state. 

It is convenient to start from a $[U(N)\times{\cal{M}}]$-invariant gapless point in the gauge choice (\ref{M}): 
\begin{eqnarray}
\label{P/Q}
H_{0,p/q}=\sum_{\vec{r}}c^\dagger_{\vec{r}+\hat{x}}c_{\vec{r}}+c^\dagger_{\vec{r}+N_c\hat{y}}c_{\vec{r}}(-1)^{r_x}+\text{h.c.}, 
\end{eqnarray}
which will be gapped later by arbitrary interactions respecting $[U(N)\times{\cal{M}}]$ symmetry. The lattice model $H_{0,p/q}$ possesses an exact anti-unitary time-reversal (TR) symmetry $\mathbb{Z}_2^T$: $c_{\vec{r}}\rightarrow c_{\vec{r}}$ with $i\rightarrow-i$ and $\mathbb{Z}_2^T$ will play a similar role as ${\cal{C}}$ in the $(0+1)$-dimensional case. 

The low-energy effective theory of (\ref{P/Q}) consists of $2$ valleys, $N_c$ colors and $N$ flavors of massless complex Dirac fermion on a $(2+1)$-dimensional manifold $X_3$: 
\begin{eqnarray}
\label{dirac}
{\cal{L}}_{0}(\bar{\psi},\psi,{\cal{A}},\omega)\!=\!\!\sum_{v=1}^2\sum_{c=1}^{N_c}\sum_{f=1}^Ni\bar{\psi}_f^{(v,l)}\left(i{\cal{D}}_3\right)^{ff'}_{(v,l);(v',l')}\psi_{f'}^{(v',l')}, \nonumber\\
\end{eqnarray}
where ${\cal{D}}_3\equiv\sum_{j=0}^2\gamma^j(\partial_j+\omega_j-i{\cal{A}}_j)$ is the three-dimensional Dirac operator with $\{\gamma^j,\gamma^k\}=-2g^{jk}$, $g^{jk}$ the metric of the manifold $X_3$ with $\omega_j$ the spin connection~\cite{Eguchi:1980aa,Nakahara:2003aa,Green:2012aa}, $\bar{\psi}$ the adjoint of $\psi$ and ${\cal{A}}=A_{U(N)}\otimes\mathbb{I}_{{\cal{M}}}+\mathbb{I}_{U(N)}\otimes A_{{\cal{M}}}$ the [$U(N)\times{\cal{M}}$]-connection, where $\mathbb{I}_{U(N)}$ and $\mathbb{I}_{{\cal{M}}}$ are identities in $u(N)$- and ${\cal{M}}$-spaces. 
The valley indices $(c=1,2)$ come from two gap-closing points in the momentum space. 
$N_c$ colors result from that (\ref{P/Q}) is separable into $N_c$ of decoupled subsystems. 
The magnetic translational symmetry ${\cal{M}}$ acts on the valley indices $v$ and the color indices $c$ so its connection $A_{\cal{M}}$ carries both valley and color indices. 
%TR symmetry transforms ${\cal{A}}$ covariantly so that the Dirac equation is TR-invariant. 
The $2N_cN=Nq$ components of Dirac fermions constitute an irreducible representation of $U(N)\times{\cal{M}}$, which implies the [$U(N)\times{\cal{M}}$]-symmetric P-V regulator $\chi$ must have a diagonal mass term: ${\cal{L}}_{0,\text{P-V}}={\cal L}_0+i\bar{\chi}(i{\cal{D}}_3-\mu^{\,}_\text{P-V})\chi$, where the summation notations of valley, color and flavor indices are suppressed. 

However, the P-V regulator mass term or any other [$U(N)\times{\cal{M}}$]-symmetric local regulator inevitably breaks $\mathbb{Z}_2^T$ due to a mixed [$(U(N)\times{\cal{M}})\rtimes\mathbb{Z}_2^T$] anomaly~\cite{Yao:2019aa}. 
According to APS index theorem and the fact that $\mathbb{Z}_2^T$ is respected by the lattice gapless point (\ref{P/Q}), we need to regularize the partition function of the low-energy effective theory (\ref{dirac}) by a bulk term defined on $Y_4$ with $\partial Y_4=X_3$ so that [$(U(N)\times{\cal{M}})\rtimes\mathbb{Z}_2^T$] is restored: 
\begin{eqnarray}
\label{bulk_4}
Z_{0}[{\cal{A}},\omega]&=&Z_{0,\text{P-V}}[{\cal{A}},\omega]\nonumber\\
&&\cdot\exp\left\{\int_{Y_4}i\pi\left[\frac{\text{Tr}{\cal{F}}^2}{8\pi^2}+\frac{Nq}{48}\frac{\text{tr}{\cal{R}}^2}{(2\pi)^2}\right]\right\}, 
\end{eqnarray}
where ${\cal{F}}$ and ${\cal{R}}$ are the $2$-form curvature tensors of connections ${\cal{A}}$ and $\omega$, respectively~\cite{Eguchi:1980aa,Nakahara:2003aa,Green:2012aa}, ``Tr'' is the trace on the valley, color and flavor indices while ``tr'' is taken on spacetime indices. The products between differential forms are understood as wedge products. 

Similarly to the $(0+1)$-dimensional case, we apply an equivalent bulk-interface-bulk formulation: a bulk Dirac fermion is defined on an extended bulk $\tilde{Y}_4$ constructed by pasting $Y_4$ and the vacuum $Y_{4,\text{vac}}$ along their interface $X_3=\partial Y_4=-\partial Y_{4,\text{vac}}$, where the bulk Dirac mass, deeply in $Y_4$, has an opposite sign to the bulk P-V regulator mass and has the same sign as the bulk P-V regulator mass deeply in the vacuum $Y_{4,\text{vac}}$~\cite{Suppl}. Our gapless degrees of freedom (\ref{P/Q}) are on the interface between the vacuum $Y_{4,\text{vac}}$ and $Y_4$, where the bulk Dirac mass vanishes~\cite{Jackiw:1976aa}. 
The partition function of the bulk Dirac field on $\tilde{Y}_4$ is equal to (\ref{bulk_4}) modulo a non-universal factor~\cite{Suppl}. 
Thus gapping this gapless point by a local interaction in $(2+1)$ dimensions is equivalent to gapping the bulk-interface-bulk system by local interface interactions. 
Integrating out the matter fields on $\tilde{Y}_4$, we calculate the resultant partition function and it contains the bulk term which is only deformed near the interface due to the local interface interaction~\cite{Suppl}: 
\begin{eqnarray}
\label{resp}
Z[{\cal{A}},\omega]=\exp\left\{\int_{\tilde{Y}_4}i\theta_1(x)\left[\frac{\text{Tr}{\cal{F}}^2}{8\pi^2}+\frac{Nq}{48}\frac{\text{tr}{\cal{R}}^2}{(2\pi)^2}\right]\right\}, \nonumber\\
\end{eqnarray}
where $\theta_1(x)=\pi$ deeply in $Y_4$ while $\theta_1(x)=\theta_\text{vac}=-2\pi n_1$ deeply in $Y_{4,\text{vac}}$ with $n_1\in\mathbb{Z}$. $\theta_\text{vac}$ is the theta-angle identified with that of the vacuum~\cite{theta}. 

%and the reason that both the electromagnetic and the gravitational theta-angles are rotated by the same $\theta_1(x)$ is that they are simultaneously generated by a global chiral anomaly phase factor of a massive Dirac fermion in $(3+1)$ dimensions~\cite{Fujikawa:2004aa,Fujikawa:2004ab,Seiberg:2016aa,Yao:2019aa} according to the interface formulation of the bulk regularization~\cite{Suppl}, which implies that we can imagine our system $X_3$ as the interface of the vacuum and $Y_4$. Here $\theta(x)=\pi$ deeply in $Y_4$ while $\theta(x)=\theta_\text{vac}$ deeply in $Y_{4,\text{vac}}$ where $\theta_\text{vac}$ is the theta-angle identified with the vanishing theta-angle of the vacuum~\cite{theta}.
Since the vacuum can be assigned a zero theta-angle, $\theta_\text{vac}$ is the period of theta-angle. Thus it can be calculated in an alternative way as follows. $\theta_\text{vac}$ is determined by the quantization property of the exponential of (\ref{bulk_4}) for any closed manifold $\hat{Y}_4$ realizable by a time-dependent lattice specified by ${\cal{A}}$. 
We can prove that
\begin{eqnarray}
\label{quant}
\int_{\hat{Y}_4}\frac{\text{Tr}{\cal{F}}^2}{8\pi^2}+\frac{Nq}{48}\frac{\text{tr}{\cal{R}}^2}{(2\pi)^2}\in\mathbb{Z}. 
\end{eqnarray}
The quantization condition (\ref{quant}) gives the same result $\theta_\text{vac}=-2\pi n_1$ with $n_1\in\mathbb{Z}$ since the effective action is defined modulo $2\pi i$. 

The construction of $\hat{Y}_4$ and ${\cal{A}}$ for the minimal value $\pm1$ of the right-hand side of (\ref{quant}) above can be done in the following way. We set $\hat{Y}_4=T^4$ a four-dimensional torus which can be realized by a spacetime lattice. Then unit fluxes of ${\cal{M}}_x$ and ${\cal{M}}_y$ are inserted through the $S^1$ components of $\tau$- and $x$-directions, respectively. Furthermore, we insert a fractional $2\pi (p/q)\mathbb{I}_{U(N)}$ flux into the $T^2$ spanned by $\tau$-$x$ plane and a $2\pi(\mathbb{I}_{U(N)}-\lambda_{N^2-1})/N$ flux into the $T^2$ spanned by $y$-direction and the extra fourth dimension, where $\lambda_{N^2-1}=\text{diag}[1,1,\cdots,1,-(N-1)]$ is the analog of the last Cartan matrix in the Gell-Mann basis~\cite{Georgi:2018aa}. 
Such a gauge configuration has been shown to exist on a lattice with a tilted boundary condition~\cite{Yao:2019aa}, 
\begin{eqnarray}
c_{\vec{r}+L_x\hat{x}}=c_{\vec{r}+\hat{y}};\,\,c_{\vec{r}+L_y\hat{y}}=c_{\vec{r}}, 
\end{eqnarray}
where $L_{x,y}$ is the loop length in $x$- and $y$-directions. It should be noted that the fractional flux inserted through $\tau$-$x$ plane exactly cancels the ambiguity induced by the tilted boundary conditions along $\tau$- and $x$-direction. 
%Thus ${\cal{A}}$ is {non-separable} into globally well-defined ${\cal{A}}_{U(N)}$ and ${\cal{A}}_{{\cal{M}}}$. 
A mathematical formulation of the bundle above can be found in \cite{Suppl}.

\section{Constraints on Hall conductances and a Wiedemann-Franz type law}
Integrating (\ref{resp}) by part, we obtain
\begin{eqnarray}
\label{C-S}
Z[{\cal{A}},\omega]&=&\exp\left[i\frac{(n_1+1/2)}{4\pi}\int_{X_3}\text{Tr}\left({\cal{A}}d{\cal{A}}+\frac{2}{3}{\cal{A}}^3\right)\right]\nonumber\\
&&\cdot\exp\left[i\frac{Nq(n_1+1/2)}{96\pi}\int_{X_3}\text{tr}\left(\omega d\omega+\frac{2}{3}\omega^3\right)\right], \nonumber\\
\end{eqnarray}
which is independent of the bulk extension $Y_4$ and the field strength defined on $Y_4$, as expected by the fact that the lattice theory is purely $(2+1)$-dimensional. 
To obtain the $U(1)$ electric response, we take ${\cal{A}}=A_{U(1)}\mathbb{I}_{U(N)}\otimes\mathbb{I}_{{\cal{M}}}$ and the ``Tr'' is done trivially giving $Nq$ in (\ref{C-S}). The levels of the resultant C-S terms above indicate, in the low-temperature limit~\cite{Luttinger:1964aa,Stone:2012aa,Nakai:2017aa}: 
\begin{eqnarray} 
\label{w-f}
\frac{\kappa_{xy}}{\sigma_{xy}}=1, 
\end{eqnarray}
and both $\kappa_{xy}$ and $\sigma_{xy}$ are constrained by: 
\begin{eqnarray}
\label{hall}
\kappa_{xy}=\sigma_{xy}=Nq/2\mod Nq, 
\end{eqnarray}
since $n_1\in\mathbb{Z}$. 
The resultant W-F law (\ref{w-f}) also directly implies that we can obtain both Hall conductances by measuring either of them, which is useful in real experiments. 
In addition, the nontrivial ``$Nq/2$'' factor in (\ref{hall}) purely results from the non-local bulk regulator in (\ref{bulk_4}) at the gapless point while any local $[U(N)\times{\cal{M}}]$-symmetric regulator, e.g. P-V regulators, can only contribute to the ``mod $Nq$'' part rather than the $Nq$-half part. 
%Thus we cannot obtain the correct conductances without the bulk regularization. 

\section{Observable consequences and constraints on gapped phases}
%Our main results are constituted with (\ref{w-f}) and (\ref{hall}) and they have significant constraints on the possible symmetric gapped phases with a unique ground state. 
The constraint (\ref{hall}) implies that both electric and thermal Hall conductances cannot vanish. Therefore, there are always charged chiral edge modes of a finite half-filled IQH material as long as [$U(N)\times{\cal{M}}$] is respected. 

The W-F law (\ref{w-f}) means that we cannot form a symmetric cluston Chern insulator by binding odd electrons together to a Chern band~\cite{Wang:2014aa,Wang:2014ab,Wang:2014ac}.
A pure thermal Hall insulator, such as a symmetric Mott insulator formed by $SU(N)$-fundamental flavors, is not permitted, either. 
This consequence is related to the generalized LSM theorem of $SU(N)$ spin system~\cite{Yao:2019aa,Yao:2019ab} that $SU(N)$ Young-tableau boxes per unit cell being indivisible by $N$ excludes the gapped phase with a unique ground state~\footnote{the magnetic translation ${\cal{M}}$ is reduced to the conventional lattice translational symmetry for $SU(N)$ generators $S^\alpha_\beta$ since charges are frozen in the low-energy limit by the parton relation $S^\alpha_\beta=\left(c^\dagger_\alpha c_\beta-\delta_{\alpha\beta}\right)$ which is charge neutral~\cite{Affleck:1987aa,Affleck:1988aa}. }.
%, where the magnetic translation ${\cal{M}}$ is reduced to the conventional lattice translational symmetry for $SU(N)$ generators $S^\alpha_\beta$ since charges are frozen in the low-energy limit by the parton relation $S^\alpha_\beta=\left(c^\dagger_\alpha c_\beta-\delta_{\alpha\beta}\right)$ which is charge neutral~\cite{Affleck:1987aa,Affleck:1988aa}. 

\section{Conclusions and discussions}
We proposed the necessity and significance of non-local bulk regularizations, which are essential to obtain correct electric and thermal responses. 
The thermal and electric IQH conductances obey a W-F law and they are constrained as $Nq/2\mod Nq$ in $N$-flavor interacting fermionic system on any half-filled two-dimensional lattice with a rational magnetic flux $2\pi p/q$ per plaquette in the presence of $U(N)$ and magnetic translational symmetries. 
Our results also restrict the permitted gapped phases. 

The constraint of the electric IQH conductance in $N=1$ case has been obtained by earlier works~\cite{Zak:1964aa,Zak:1964ab,Lu:2017aa}, but the thermal Hall conductance and W-F law can be hardly seen or investigated in their lattice approach. 
In contrast, our work treats both thermal and electric Hall conductances in a unified way with a general $U(N)$ onsite symmetry, which has potential applications for other symmetries. 
Our method also generalizes the time-reversal broken surfaces of (3+1)-dimensional topological insulators (TIs), which possess half-integer thermal and electric Hall conductances on surfaces~\cite{Qi:2008aa,Qi:2011aa,Wang:2014aa,Wang:2014ab}. 

Furthermore, the generalizations of the onsite symmetry to other continuous symmetries and higher dimensions~\cite{Florek:1996aa} can be expected. The constraints on fractional quantum Hall effects will be future interest. 

\section{Acknowledgements}
I am grateful to Yuji Tachikawa for the constructive advice on the $(0+1)$-dimensional case and higher dimensions, and I am thankful to Jyong-Hao Chen, Meng Cheng, Chang-Tse Hsieh, Shiyong Liu, Masaki Oshikawa, Shinsei Ryu and Yasuhiro Tada for the useful discussions. 
I was supported by JSPS fellowship.
This work was supported in part by JSPS KAKENHI Grant No. JP19J13783 and the National Science Foundation under Grant No. NSF PHY-1748958.

\sloppy
\bibliography{bib}
\clearpage
\appendix*
\begin{widetext}
\section{An equivalence between the bulk regularization and the interface formulation}
In this part, we will show that the partition function of a massless Dirac fermion on $X_{2k+1}$ of odd dimension(s) $(2k+1)$ under bulk regularization by $Y_{2k+2}$ on which the gauge structures on $X$ is extended
\begin{eqnarray}
\label{full}
Z[{\cal{A}},\omega]=Z_\text{P-V}[{\cal{A}},\omega]\exp\left(\int_{Y_{2k+2}:\partial Y_{2k+2}=X_{2k+1}} i\pi\hat{A}({\cal{R}})\text{ch}({\cal{F}})\right)
\end{eqnarray}
is, modulo a non-universal coefficient, equivalent to the partition function of a massive Dirac fermion on $\tilde{Y}_{2k+2}$ where $\tilde{Y}_{2k+2}$ is constructed by pasting $Y_{2k+2}$ with $Y_{2k+2,\text{vac}}$ along their common boundary $X_{2k+1}$ called an ``interface'' and there is a local structure of $\tilde{Y}_{2k+2}$ near the pasted boundary $X_{2k+1}$ as $X_{2k+1}\times\mathbb{R}$ with $X_{2k+1}\times\{0\}$ identified with $X_{2k+1}=\partial Y_{2k+2}=-\partial Y_{2k+2,\text{vac}}$ and a flat $\mathbb{R}$. $\hat{A}({\cal{R}})$ is the genus and $\text{ch}({\cal{F}})$ is the total Chern character. 
Here the integral is understood to be done on the volume form of the integrated form. Let us parametrize this extra dimension by $\mathbb{R}: s\in(-\infty,+\infty)$ and $X_{2k+1}\times\mathbb{R}_-$ is in $Y_{2k+2,\text{vac}}$ and we take $\left\{\gamma^{\mu},\gamma^\nu\right\}=-2g^{\mu\nu}$ including the index $s$. 
We consider the case that the Dirac spinor forms an irreducible representation of the gauge group, so the P-V regulator mass is diagonal in the gauge-symmetry space. 

The (real) mass of the Dirac fermion on $X_{2k+1}\times\mathbb{R}$ varies as
\begin{eqnarray}
m(s)=\left\{\begin{array}{cc}-\mu,&s\in(-\infty,-l];\\+\mu,&s\in[+l,+\infty)\end{array}\right., 
\end{eqnarray}
where $\mu$ is the opposite value of the P-V regulator mass on $\tilde{Y}_{2k+2}$ to be defined below and $l$ is a non-universal length scale and $m(s)$ must have at least one zero in $s\in(-l,l)$ due to being real. Thus the Lagrangian of the massive fermion $\Psi$ is
\begin{eqnarray}
\label{tildeY}
{\cal{L}}_{\tilde{Y}_{2k+2}}=i\bar{\Psi} \left[i{\cal{D}}_{\tilde{Y}_{2k+2}}+m(s)\right]\Psi, 
\end{eqnarray}
and its regulator
\begin{eqnarray}
\label{regtildeY}
{\cal{L}}_\text{reg}=i\bar{\chi} \left[i{\cal{D}}_{\tilde{Y}_{2k+2}}-\mu\right]\chi. 
\end{eqnarray}

\subsection{Localized excitations on the interface}
We first generate the massless modes $Z_\text{P-V}[{\cal{A}},\omega]$ in (\ref{full})
Let us consider the boundary mode ansatz
\begin{eqnarray}
\Psi_{X_{2k+1}}(s)=\psi\exp[-m(s)s], 
\end{eqnarray}
which is localized within the interface $(-l,+l)$ and is ``chiral'': $(1-i\gamma^s)\psi=0$. 

We insert the ansatz into the Lagrangian and obtain
\begin{eqnarray}
\label{local}
{\cal{L}}_{\tilde{Y}_{2k+2}}|_\text{boundary}&=&i\bar{\Psi}_{X_{2k+1}}(s)\gamma^s[\partial_s+i\bar{\gamma}{\cal{D}}_{X_{2k+1}}+i\gamma^s m(s)]\Psi_{X_{2k+1}}(s)\nonumber\\
&=&i\bar{\psi}(i{\cal{D}}_{X_{2k+1}})\psi+{\cal{O}}(l), 
\end{eqnarray}
where $\bar{\gamma}\equiv\gamma^s\gamma^1\cdots$ the (hermitian) chirality operator with $\bar{\gamma}^2=1$ and ${\cal{O}}(l)$ a non-universal part depending on $l$. Here the Dirac operator ${\cal{D}}_{X_{2k+1}}$ on $X_{2k+1}$ is expressed in the bulk terms restricted on $X_{2k+1}\times\{0\}$: 
\begin{eqnarray}
{\cal{D}}_{X_{2k+1}}=\sum_{\mu\neq s}-\bar{\gamma}\gamma^s\gamma^\mu D_\mu. 
\end{eqnarray}
It is clear that the localized modes obtained in (\ref{local}) is exactly the massless Dirac fermion on $X_{2k+1}$. Its partition function, after regularized by the P-V regulator, is exactly $Z_\text{P-V}[{\cal{A}},\omega]$ in (\ref{full}). 

\subsection{The bulk responses}
Since we have obtained the localized modes within $X_{2k+1}\times(-l,+l)$, we can take $l$ to be small and omit the (localized) excitation there. Then we do the following chiral transformation on $\Psi$: 
\begin{eqnarray}
\Psi'=\exp[i\bar{\gamma}\varphi(s)/2]\Psi
\end{eqnarray}
with
\begin{eqnarray}
\varphi(s)=\left\{\begin{array}{cc}-2\pi k,&s\in(-\infty,-l]\\\pi,&s\in[+l,+\infty)\end{array}\right., 
\end{eqnarray}
where $k\in\mathbb{Z}$. By this chiral transformation, beyond the interface $(-l,+l)$, the form of the bulk Lagrangian is precisely the same as its regulator (\ref{regtildeY}). Therefore, by Fujikawa's method, the additional term induced by a global chiral anomaly is
\begin{eqnarray}
Z_\text{bulk}[{\cal{A}},\omega]=\exp\left[\int_{Y_{2k+2}:\partial Y_{2k+2}=X_{2k+1}} i(2k+1)\pi\hat{A}({\cal{R}})\text{ch}({\cal{F}})+{\cal{O}}(l)\right]. 
\end{eqnarray}
Combined with (\ref{local}), the full partition function is, modulo non-universal factors, exactly (\ref{full}) after $k=0$ is taken due to the explicit time-reversal symmetry of (\ref{tildeY}) together with its regulator (\ref{regtildeY}) by the APS index theorem. However, if without the time-reversal symmetry, any other $k\in\mathbb{Z}$ is possible which corresponds to, after the interface is gapped, deposit of integer quantum Hall of level $k$ on the interface thereby breaking time reversal explicitly. 

Then combining the gapless interface mode and the bulk effective theory with $k=0$ required by time reversal, we obtain the partition function of the Lagrangian (\ref{tildeY})
\begin{eqnarray}
Z_{\tilde{Y}_{2k+2}}[{\cal{A}},\omega]&=&Z_\text{P-V}[{\cal{A}},\omega]Z_\text{bulk}[{\cal{A}},\omega]\exp[{\cal{O}}(l)]\nonumber\\
&=&Z_\text{P-V}[{\cal{A}},\omega]\exp\left[\int_{Y_{2k+2}:\partial Y_{2k+2}=X_{2k+1}} i\pi\hat{A}({\cal{R}})\text{ch}({\cal{F}})+{\cal{O}}(l)\right]\nonumber\\
&=&Z[{\cal{A}},\omega]\exp[{\cal{O}}(l)]. 
\end{eqnarray}
Therefore, we have shown the physical equivalence of the bulk regularization (\ref{full}) and the interface formulation (\ref{tildeY}) between the vacuum and a potentially nontrivial bulk in the presence of some certain symmetry. 

\section{The gapped interface by local time-reversal breaking interactions in $(2+1)$ dimensions}
We have show the equivalence of the bulk regularization scheme with the bulk-interface-bulk formulation. Thus gapping the original system in $(2+1)$ dimensions is equivalent to gapping the interface by interactions acted on the interface. Since the bulks on both sides of the interface is gapped, it is further the same as gapping the whole bulk-interface-bulk system by an interface interaction. Therefore, the responses deep in two bulks is expected unchanged by the local interface interaction as we will see. 

The relevant interaction allowed by Lorentz invariance is the gauge-diagonal chiral and the Dirac mass terms in $(2k+2)$ dimensions but only non-vanishing locally around the interface in $(2k+1)$ dimensions: 
\begin{eqnarray}
{\cal{L}}_{\tilde{Y}_{2k+2},\text{gapped}}=i\bar{\Psi} \left[i{\cal{D}}_{\tilde{Y}_{2k+2}}+\mu\exp[i\bar{\gamma}\theta_k(s)]\right]\Psi, 
\end{eqnarray}
where we have normalized the mass term without loss of generality in the gapped phase and
\begin{eqnarray}
\theta_k(s)=\left\{\begin{array}{ll}\pi\mod2\pi,&s\rightarrow-\infty;\\0\mod2\pi,&s\rightarrow+\infty\end{array}\right., 
\end{eqnarray}
due to the locality of the interface interaction. The regulator still takes the form as (\ref{regtildeY}). 
Therefore, 
\begin{eqnarray}
\theta_k(+\infty)-\theta_k(-\infty)\equiv(2n_{k}+1)\pi, 
\end{eqnarray}
with $n_k\in\mathbb{Z}$. 
To calculate the partition function, we take the following chiral transformation
\begin{eqnarray}
\Psi''=\exp[i\bar{\gamma}\theta_k(s)/2]\Psi, 
\end{eqnarray}
to eliminate the chiral phase of the mass term in the price of a response term in the partition function by the chiral anomaly: 
\begin{eqnarray}
Z_{\tilde{Y}_{2k+2},\text{gapped}}[{\cal{A}},\omega]&=&\int{\cal{D}}[\bar{\Psi},\Psi]\exp\left(-\int_{\tilde{Y}_{2k+2}}{\cal{L}}_{\tilde{Y}_{2k+2},\text{gapped}}+{\cal{L}}_\text{reg}\right)\nonumber\\
&=&\exp\left[\int_{\tilde{Y}_{2k+2}}i\theta_k(s)\hat{A}({\cal{R}})\text{ch}({\cal{F}})\right]. 
\end{eqnarray}
By integrations by part, we obtain
\begin{itemize}
\item $k=0$: 
\begin{eqnarray}
Z_{\tilde{Y}_{2},\text{gapped}}[{\cal{A}}]=\exp\left[\int_{X_{1}}i(n_0+1/2)\text{Tr}({\cal{A}})\right],  
\end{eqnarray}
and gravitational responses are irrelevant in this dimension. 
\item $k=1$: 
\begin{eqnarray}
Z_{\tilde{Y}_{2},\text{gapped}}[{\cal{A}},\omega]&=&\exp\left[\frac{i(n_1+1/2)}{4\pi}\int_{X_3}\text{Tr}\left({\cal{A}}d{\cal{A}}+\frac{2}{3}{\cal{A}}^3\right)\right]\nonumber\\
&&\cdot\exp\left[\frac{i\text{Tr}(1)(n_1+1/2)}{96\pi}\int_{X_3}\text{tr}\left(\omega d\omega+\frac{2}{3}\omega^3\right)\right],
\end{eqnarray}
where $\text{Tr}(1)$ is the dimension(s) of the gauge bundle with connection ${\cal{A}}$. 
\end{itemize}
The results above exactly reproduce the various responses in the main text. 

\section{Gauge bundle for the quantized theta-term with a unit instanton in $(3+1)$ dimensions}
Let us give a mathematically rigorous construction of the gauge bundle of all the symmetries. We first re-parametrize the torus $T^4$ by rescaling its four components as $L_{\tau,x,y,z}\rightarrow2\pi$, and then $\tau,x,y,z\in\mathbb{R}/2\pi$ with $\mathbb{R}$ the universal covering of each $S^1$-component of $T^4$. The symmetry-group gauge bundle can be expressed by transition function between various patches. In order to do so, we need to cover the $T^4$ by four open areas: 
\begin{eqnarray}
\label{gauge_bundle}
&&\left\{\begin{array}{l}{\mathscr{A}}_\text{I}=\{p\in T^4|\phi_\text{I}(p)=(\tau_\text{I},x_\text{I},y_\text{I},z_\text{I})\in(-\epsilon,\pi+\epsilon)\times(-\epsilon,\pi+\epsilon)\times[0,2\pi)\times(-\epsilon,\pi+\epsilon)\},\\
{\mathscr{A}}_\text{II}=\{p\in T^4|\phi_\text{II}(p)=(\tau_\text{II},x_\text{II},y_\text{II},z_\text{II})\in(-\epsilon,\pi+\epsilon)\times(-\pi-\epsilon,+\epsilon)\times[0,2\pi)\times(-\epsilon,\pi+\epsilon)\},\\
{\mathscr{A}}_\text{III}=\{p\in T^4|\phi_\text{III}(p)=(\tau_\text{III},x_\text{III},y_\text{III},z_\text{III})\in(-\pi-\epsilon,\epsilon)\times(-\pi-\epsilon,\epsilon)\times[0,2\pi)\times(-\epsilon,\pi+\epsilon)\},\\
{\mathscr{A}}_\text{IV}=\{p\in T^4|\phi_\text{IV}(p)=(\tau_\text{IV},x_\text{IV},y_\text{I},z_\text{IV})\in(-\pi-\epsilon,+\epsilon)\times(-\epsilon,\pi+\epsilon)\times[0,2\pi)\times(-\epsilon,\pi+\epsilon)\},\end{array}\right.\nonumber\\
&&\left\{\begin{array}{l}{\mathscr{A}}'_\text{I}=\{p\in T^4|\phi'_\text{I}(p)=(\tau'_\text{I},x'_\text{I},y'_\text{I},z'_\text{I})\in(-\epsilon,\pi+\epsilon)\times(-\epsilon,\pi+\epsilon)\times[0,2\pi)\times(-\pi-\epsilon,\epsilon)\},\\
{\mathscr{A}}'_\text{II}=\{p\in T^4|\phi'_\text{II}(p)=(\tau'_\text{II},x'_\text{II},y'_\text{II},z'_\text{II})\in(-\epsilon,\pi+\epsilon)\times(-\pi-\epsilon,+\epsilon)\times[0,2\pi)\times(-\pi-\epsilon,\epsilon)\},\\
{\mathscr{A}}'_\text{III}=\{p\in T^4|\phi'_\text{III}(p)=(\tau'_\text{III},x'_\text{III},y'_\text{III},z'_\text{III})\in(-\pi-\epsilon,\epsilon)\times(-\pi-\epsilon,\epsilon)\times[0,2\pi)\times(-\pi-\epsilon,\epsilon)\},\\
{\mathscr{A}}'_\text{IV}=\{p\in T^4|\phi'_\text{IV}(p)=(\tau'_\text{IV},x'_\text{IV},y'_\text{I},z'_\text{IV})\in(-\pi-\epsilon,+\epsilon)\times(-\epsilon,\pi+\epsilon)\times[0,2\pi)\times(-\pi-\epsilon,\epsilon)\},\end{array}\right.\nonumber\\
\end{eqnarray}
where $\epsilon$ is a smaller positive number than $\pi/2$, e.g. $\epsilon=\pi/4$, and we have used the coordinates $\phi_\text{I,II,III,IV}(p\in T^4)$ of the universal covering $\mathbb{R}^4$ to coordinate $T^4$. 

%For simplicity, we define $U_{\sqrt{\eta_{1,2}}}$ as the $U(1)_Q$ transformation creating a phase of ($\sqrt{\eta_{1,2}}$) for a single fermion operator $\bar{\psi}$. 
We denote the transition function as, e.g. $\psi_\text{I}=t_\text{I,II}\psi_\text{II}$ and \emph{etc.}, and the following transition functions are defined: 

($\lambda_0$ denotes the identity matrix and $\lambda_{N^2-1}=\text{diag}[1,1,\cdots,1,-(N-1)]$)
\begin{itemize}
\item Transition function between ${\mathscr{A}}_\text{I}$ and ${\mathscr{A}}_\text{II}$ on ${\mathscr{A}}_\text{I}\cap{\mathscr{A}}_\text{II}$: 
\begin{eqnarray}
\label{trans_1_2}
t_\text{I,II}\circ\phi^{-1}_\text{I}(\tau_\text{I},x_\text{I},y_\text{I},z_\text{I})=\left\{\begin{array}{ll}T_2,&(x_\text{I},\tau_\text{I})\in(-\epsilon,+\epsilon)\times(-\epsilon,\pi+\epsilon),
\\
1,&(x_\text{I},\tau_\text{I})\in(\pi-\epsilon,\pi+\epsilon)\times(-\epsilon,\pi+\epsilon), 
\end{array}\right.
\end{eqnarray}
where $\tau_\text{I}$ and $x_\text{I}$ are free parameters in their own domains. 
\item Transition function between ${\mathscr{A}}_\text{IV}$ and ${\mathscr{A}}_\text{III}$ on ${\mathscr{A}}_\text{IV}\cap{\mathscr{A}}_\text{III}$: 
\begin{eqnarray}
\label{trans_4_3}
t_\text{IV,III}\circ\phi^{-1}_\text{IV}(\tau_\text{IV},x_\text{IV},y_\text{IV},z_\text{IV})=\left\{\begin{array}{ll}T_2,&(x_\text{IV},\tau_\text{IV})\in(-\epsilon,+\epsilon)\times(-\pi-\epsilon,\epsilon),
\\
1,&(x_\text{IV},\tau_\text{IV})\in(\pi-\epsilon,\pi+\epsilon)\times(-\pi-\epsilon,\epsilon). 
\end{array}\right.
\end{eqnarray}
\item Transition function between ${\mathscr{A}}_\text{II}$ and ${\mathscr{A}}_\text{III}$ on ${\mathscr{A}}_\text{II}\cap{\mathscr{A}}_\text{III}$: 
\begin{eqnarray}
t_\text{II,III}\circ\phi^{-1}_\text{II}(\tau_\text{II},x_\text{II},y_\text{II},z_\text{II})=\left\{\begin{array}{ll}-T_1\exp\left(- i {\lambda_0}\frac{p}{q}x_\text{II}\right),&(x_\text{II},\tau_\text{II})\in(-\pi-\epsilon,\epsilon)\times(-\epsilon,\epsilon),
\\
1,&(x_\text{II},\tau_\text{II})\in(-\pi-\epsilon,\epsilon)\times(\pi-\epsilon,\pi+\epsilon). 
\end{array}\right.
\end{eqnarray}
\item Transition function between ${\mathscr{A}}_\text{I}$ and ${\mathscr{A}}_\text{IV}$ on ${\mathscr{A}}_\text{I}\cap{\mathscr{A}}_\text{IV}$: 
\begin{eqnarray}
t_\text{I,IV}\circ\phi^{-1}_\text{I}(\tau_\text{I},x_\text{I},y_\text{I},z_\text{I})=\left\{\begin{array}{ll}-T_1\exp\left(- i {\lambda_0}\frac{p}{q}x_\text{I}\right),&(x_\text{I},\tau_\text{I})\in(-\epsilon,\pi+\epsilon)\times(-\epsilon,\epsilon),
\\
1,&(x_\text{I},\tau_\text{I})\in(-\epsilon,\pi+\epsilon)\times(\pi-\epsilon,\pi+\epsilon). 
\end{array}\right.
\end{eqnarray}
\item Transition functions between ${\mathscr{A}}'_{i}$ and ${\mathscr{A}}'_j$ on ${\mathscr{A}}'_i\cap{\mathscr{A}}'_j$ are naturally identified to that of $t_{i,j}$ through $t'_{i,j}\circ{\phi'_{i}}^{-1}(\tau',x',y',z')=t_{ij}\circ\phi_{i}^{-1}(\tau',x',y',z'+\pi)$. 
\item Transition functions between ${\mathscr{A}}_i$ and ${\mathscr{A}}_i'$ on ${\mathscr{A}}_i\cap{\mathscr{A}}_i'$: 
\begin{eqnarray}
t_{i,i'}\circ\phi^{-1}_i(\tau_{i},x_{i},y_i,z_i)=\left\{\begin{array}{ll}\exp\left(- i \frac{\lambda_0-\lambda_{N^2-1}}{N}y_i\right),&z_i\in(-\epsilon,\epsilon),
\\
1,&z_i\in(\pi-\epsilon,\pi+\epsilon). 
\end{array}\right.
\end{eqnarray}
\end{itemize}

Then
\begin{eqnarray}
%\label{quant}
\int_{\hat{Y}_4=T^4}\frac{\text{Tr}{\cal{F}}^2}{8\pi^2}+\frac{Nq}{48}\frac{\text{tr}{\cal{R}}^2}{(2\pi)^2}=1. 
\end{eqnarray}

\end{widetext}

\end{document}